\documentclass[prd,aps,twocolumn,nofootinbib]{revtex4-2}

\usepackage{graphicx,psfrag}
\usepackage{mathrsfs}
\usepackage{amsmath,amsfonts,amssymb}
\usepackage{multirow}
\usepackage{comment}
\usepackage{ulem}
\usepackage{hyperref}
\usepackage{enumitem}
\usepackage{afterpage}
\usepackage{tcolorbox}
\usepackage[T1]{fontenc}
\newcommand{\be}{\begin{equation}}
\newcommand{\ee}{\end{equation}}
\newcommand{\bea}{\begin{eqnarray}}
\newcommand{\eea}{\end{eqnarray}}
\newcommand{\bel}{\begin{align}}
\newcommand{\eel}{\end{align}}

\newcommand{\mnras}{Mon.\ Not.\ R.\ Astron.\ Soc.}
\newcommand{\aap}{Astron.\ Astrophys.}

\def\GMc2{{\rm G M_{\odot} c^{-2}}}

\def\AK{{\tt AthenaK}}

\usepackage{pifont} 

\usepackage{color}
\definecolor{cyan}{rgb}{0,0.9,0.9}
\definecolor{orange}{rgb}{0.9,0.5,0}
\definecolor{magenta}{rgb}{1,0,1}
\definecolor{purple}{rgb}{0.8,0.4,0.8}
\definecolor{gray}{rgb}{0.8242,0.8242,0.8242}
\definecolor{light-gray}{gray}{0.95}

\begin{document}
\title{Magnetic field dynamics in isolated neutron stars with an external dipole field}

\author{Aurora Capobianco}
\author{William Cook}
\author{Sebastiano Bernuzzi}
\affiliation{Theoretisch-Physikalisches Institut, Friedrich-Schiller-Universit{\"a}t Jena, 07743, Jena, Germany}

\author{Brynmor Haskell}
\affiliation{Nicolaus Copernicus Astronomical Center of the Polish Academy of Sciences, Bartycka 18, 00-716, Warsaw, Poland}
\affiliation{INFN, Sezione di Milano, Via Celoria 16, 20133, Milano, Italy}
\affiliation{Dipartimento di Fisica, Università di Milano, Via Celoria 16, 20133, Italy}

\author{Jacob Fields}
\affiliation{School of Natural Sciences, Institute for Advanced Study, Princeton, NJ 08540, USA}

\date{\today}
\begin{abstract}
Neutron stars can harbor extremely strong magnetic fields, yet the structure and stability of their magnetic field configuration remain poorly understood. Observations of pulsars indicate that the large-scale external field is predominantly dipolar far from the star, while the internal magnetic configurations are largely unconstrained. We investigate the dynamical stability of magnetized neutron stars through long-term numerical-relativity simulations. We explore a range of models with an initial external dipole field and mixed poloidal-toroidal internal field where the energy of the toroidal component varies up to $90\%$ of the magnetic energy. We find that the internal magnetic field relaxes toward a dynamically stable mixed poloidal-toroidal geometry, in which the toroidal component contributes to $\lesssim10\%$ of the total magnetic energy both in the exterior and in the interior. This configuration emerges within one Alfv\'en time following the saturation of the Tayler instabilities and also aided by gravitational-wave emission.
These results suggest that long-lived neutron star magnetic fields are strongly constrained toward stable mixed configurations, with important implications for pulsar emission models, magnetar evolution, and the interpretation of gravitational-wave signals from magnetized remnants.
\end{abstract}

\maketitle

\section{Introduction}
\label{sec:intro}

Neutron stars (NSs) are compact remnants of massive stars, with magnetic fields that can exceed those of any other
observed astronomical source. The surface magnetic field of NSs is typically inferred from dipole spin-down measurements based on radio observations \citep{LorimerHandbook, Chung:2011a,Chung:2011b} and can range from $10^8\,\rm{G}$ for old-recycled pulsars to $10^{15}\, \rm{G}$ for magnetars.

Observational evidence has traditionally supported a predominantly dipolar magnetic field geometry in NSs, although indications of higher-order multipolar components have been reported \citep{deLima:2020lzl}. More recent X-ray observations, particularly from the NICER mission, have provided stronger constraints on the surface field structure through detailed hotspot modeling \citep{Gendreau:2012}. These studies suggest that the magnetic field is often far more complex than a simple aligned dipole, instead favoring configurations that include multipolar components or offset dipoles \citep{Bilous:2019,Riley:2019yda,Riley:2021pdl,Choudhury:2024,Salmi:2024}. These strong magnetic fields are thought to power the burst activity of magnetars \citep{Goldreich:1969,Thompson:1995gw}, while the internal field may be significantly stronger than the inferred external dipole, potentially inducing deformations that could result in significant gravitational wave (GW) emission \citep{Bonazzola:1995rb}. The internal field topology remains challenging to probe observationally, making theoretical modeling and numerical simulations crucial for understanding its structure.

The internal magnetic field of a NS is expected to be a complex combination of poloidal and toroidal components, as purely poloidal or purely toroidal configurations are inherently unstable to kink and varicose instabilities \citep{Tayler:1957a,Tayler:1973a,Wright:1973a,Markey:1973a,Markey:1974a,Flowers:1977a}. Dynamical simulations have shown that mixed-field configurations can evolve toward a nearly axisymmetric twisted-torus state, which is often regarded as a plausible equilibrium configuration \citep{Ciolfi:2010,Mastrano:2011,Ciolfi:2013dta,Lasky:2013,Mastrano:2015,Castillo:2017}.  This state, notably without surface currents, has a mixed geometry with the toroidal component often dominating the poloidal \citep{Braithwaite:2005xi,Braithwaite:2009} which may contribute to the long-term stability of the system.
However, the emergence of a twisted-torus geometry is not universal: numerical studies indicate that the final magnetic configuration depends sensitively on the initial radial distribution of magnetic energy \citep{Braithwaite:2008}, as well as on the relevant dissipation mechanisms and the degree of stable stratification in the stellar fluid \citep{Becerra:2022}.

Insights from similar equilibrium models in magnetized stars and white dwarfs suggest that the balance between the Lorentz force, pressure, and gravity governs these configurations, highlighting the importance of theoretical modeling and simulations for understanding the otherwise inaccessible interior field structure \citep{Braithwaite:2005md,Braithwaite:2007,Armaza:2015,Braithwaite:2008}.

Analytical studies have examined equilibrium solutions and stratification effects \citep{Chandrasekhar:1953,Tayler:1957a,Tayler:1973a,Wright:1973a,Markey:1973a,Markey:1974a,Flowers:1977a,Haskell:2007bh,Gusakov:2017uam,Ofengeim:2018zpe,Haskell:2007bh,Ciolfi:2009,Ciolfi:2011xa}. Newtonian magnetohydrodynamic (MHD) simulations have modeled the evolution and stability of poloidal-toroidal fields, including the role of twisted-torus configurations \citep{Braithwaite:2005md,Braithwaite:2005xi,Braithwaite:2007,Lander:2009ib,Lander:2011,Sur:2020hwn,Herbrik:2016,Frederick:2020}. General-relativistic magnetohydrodynamic (GRMHD) simulations have extended these studies to fully relativistic NS models, confirming the generation of toroidal components from initially poloidal fields and investigating their impact on stability and GW emission \citep{Kiuchi:2008,Ciolfi:2011xa,Lasky:2011un,Lasky:2012ju,Ciolfi:2012en,Ciolfi:2013dta,Pili:2014npa,Pili:2017yxd,Tsokaros:2021pkh,Sur:2021awe,Cheong:2024stz,Cook:2025zzy}. Together, these approaches consistently indicate that stable NS magnetic fields require a mixed poloidal-toroidal geometry, while purely poloidal or toroidal configurations are dynamically unstable \citep{Braithwaite:2005xi,Ciolfi:2009,Ciolfi:2011xa,Lander:2009ib,Tsokaros:2021pkh,Cheong:2024stz}.

Determining whether a NS’s magnetic field settles into a stable equilibrium remains a challenging and computationally demanding problem. Magnetic energy dissipation during the evolution progressively increases the Alfv\'en timescale, requiring long-duration simulations to follow the dynamics. At the same time, the growth of instabilities drives the system into a nonlinear regime where non-ideal effects and turbulence become important, requiring sufficiently high resolution to reliably capture their impact on the large-scale field structure. 

To date the longest and highest resolution simulations of isolated NSs have been performed by \citet{Cook:2025zzy} with a resolution of $29\, \rm{m}$, evolved for $100\, \rm{ms}$ and up to $1\, \rm{s}$ at lower resolution. The study employed the \AK{} code to evolve the purely poloidal magnetic field configuration at very high resolution in the Cowling approximation. In this case the generated toroidal field maintained its strength at ${\sim}15\%$ of the total magnetic field energy.

\citet{Sur:2021awe} performed GRMHD simulations with Athena++ \citep{Stone:2020}, comparing a purely poloidal field and a poloidal field superposed with a toroidal field with $80\%$ of the total magnetic field strength. They found that the purely poloidal setup generates a toroidal field which later decays exponentially reaching $1\%$ of the total magnetic energy, and the initially stronger toroidal field setup maintains a roughly constant ratio of toroidal to total magnetic energy of around $80\%$.

Both these studies focused on the dynamics and magnetic fields in the NS interior, while \citet{Sur:2020hwn} explored mixed internal configurations with an external dipole field extending from the surface with the Newtonian code PLUTO \citep{Mignone:2008ii}. They found that for both
initially predominantly poloidal and toroidal fields the field settles down to a mixed poloidal-toroidal configuration, where the toroidal component contributes between $10\%$ and $20\%$ of the total magnetic energy.

In this work, we perform nonlinear GRMHD simulations of isolated magnetized NSs with an external dipole field and explore internal mixed configurations. We study the development of the instabilities, the global evolution and energetics, and the final configuration of the magnetic field. 

In Sec. \ref{sec:method} we outline our numerical setup. Sec. \ref{sec:res} presents the results for the different configurations considered in our simulations. We begin in Section \ref{subsec:fields} by examining the qualitative evolution of the magnetic field and the associated instabilities. In Sec. \ref{subsec:eg} we analyze the energetics of the systems. In Sec. \ref{subsec:modes} we focus on the mode analysis, where we measure the radially integrated angular distribution of two key quantities: the density, $\rho$ and the radial magnetic field strength $B^r$. 
In Sec. \ref{subsec:gw} we analyze the GW emission produced by the magnetically driven oscillations of the NS. In Sec. \ref{sec:con} we summarize our conclusions and discuss the main implications of our findings and in Appendix \ref{app:conv} we elaborate on the impact of boundary effects and numerical resolution on our results.

\section{Method}
\label{sec:method}

We perform a sequence of simulations of a static magnetized NS using the code \AK{} \citep{Stone:2024,Zhu:2024utz,Fields:2024pob} in full 3D ideal GRMHD and coupling the spacetime evolution using the Z4c formulation of 3+1 Einstein equations~\citep{Bernuzzi:2009ex,Hilditch:2012fp}. The same simulations are also performed in the Cowling approximation, holding the spacetime metric fixed as that of the Tolman-Oppenheimer-Volkoff (TOV) star described below. 

For initial data we consider a non-rotating TOV star with central density $7.91\times10^{14}\; \rm{g/cm^3}$ and radius $R = 12.0\, \rm{km}$, with a Gamma law equation of state (EOS) such that $p = \rho \epsilon (\Gamma -1)$, with $\Gamma = 2$.

This is the commonly considered model A0 of \cite{Dimmelmeier:2005zk}. Outside the star, the computational domain is filled with a low density atmospheric fluid, fixed at the density $\rho_{atm} = 62\; \rm{g/cm^3}$.
We comment on the role played by the atmospheric density in Appendix \ref{app:conv}.

We place the boundaries of the computational domain at $\pm 470\, \rm{km}$ in all directions and add four static mesh refinement levels, such that the innermost level has resolution $\Delta x = 230\, \rm{m}$ and covers the NS entirely. The outer boundary of the general relativity (GR) simulations is located at ${\sim}40R$ to minimize the impact of boundary effects on our simulations. See Appendix \ref{app:conv} for a study of
outer boundary effects.
We simulate for $150\, \rm{ms}$ of evolution.

We initialize the magnetic field with the prescription from \citet{Haskell:2007bh}, where the interior is described as:
\begin{equation}
	B_r = \frac{B_p \cos \theta}{\pi(\pi^2 - 6)}[y^3 + 3(y^2 - 2)\sin y + 6y \cos y]{} 
\end{equation}
\begin{equation}
	B_\theta = \frac{B_p \sin \theta}{2\pi(\pi^2 - 6)}[-2y^3 + 3(y^2 - 2)(\sin y - y \cos y)]
\end{equation}
\begin{equation}
	B_\phi = B_t \frac{\sin y \sin \theta}{\pi}
\end{equation}

where
\begin{equation}
	y = \frac{\pi r}{R},
\end{equation}

and the exterior:
\begin{equation}
	B_r = \frac{B_p R^3 \cos\theta}{r^3}
\end{equation}

\begin{equation}
	B_\theta = \frac{B_p R^3 \sin\theta}{2 r^3}
\end{equation}

\begin{equation}
	B_\phi = 0.
\end{equation}

We therefore impose an external dipole field with maximum strength at the surface of $B_p = 10^{15}\; \rm{G}$, and a combination of toroidal and poloidal fields limited to the interior. 
The magnetic field is continuous across the surface, and axisymmetric at $t=0$. This configuration is the same as that employed by \citet{Sur:2020hwn}. While the poloidal field strength $B_p$ is fixed in all simulations, the relative toroidal strength is adjusted by increasing  $B_t$. 
We consider a set of six NS models: a non-magnetized reference model B0 and five magnetized models with initial toroidal to total magnetic field ratios ranging from $0\%$ to $90\%$ (Bt0, Bt25, Bt60, Bt70, Bt90).
No initial perturbation is imposed in our simulations; the instability instead develops from truncation errors in the Cartesian-grid discretisation.

The external dipole of strength $10^{15}\, \rm{G}$  on the surface of the star is consistent with magnetar field strength, and those previously studied by the authors in \citet{Sur:2021awe,Cook:2023bag}.
This value of $B_p$ was chosen to simulate an extreme magnetic field, while remaining sufficiently small to prevent the magnetisation on the surface from increasing excessively due to the steep density gradient. Further details are discussed in Appendix \ref{app:conv}.

\begin{figure*}[t]
	\centering
	\includegraphics[width=1\textwidth]{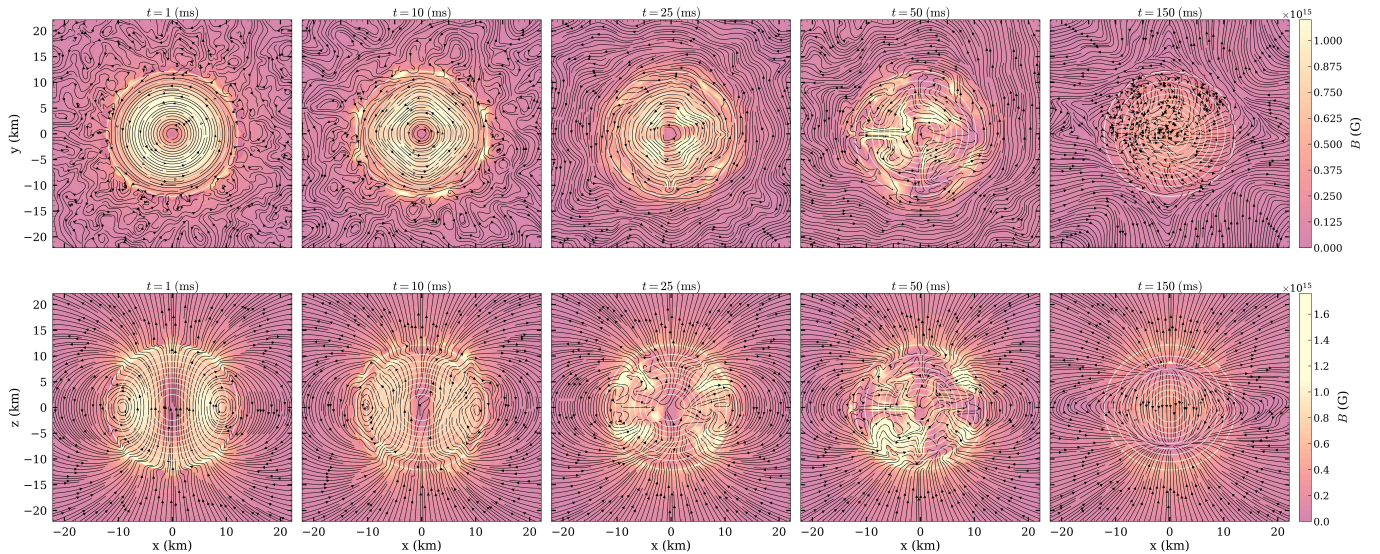}
	\caption{Snapshots from the model Bt60 of the star from meridional view (top) and equatorial view (bottom) showing the development of the magnetic field (color scale indicates the strength $B$). The black streamlines shown are the poloidal fieldlines that thread through the main body of the star. The white lines show the density contours inside the star. Times of the snapshots are given as figure titles.}
	\label{fig:2d}
\end{figure*}

\AK{} solves the coupled system of GRMHD in Eulerian conservative form and the Z4c equations. The latter provides improved constraint propagation and damping properties compared to other free-evolution schemes of Einstein equations. The GRMHD equations are solved using a finite-volume, high-resolution shock-capturing (HRSC) scheme.  Primitive variables are reconstructed at cell interfaces using the piecewise parabolic method (PPMX) \citep{Colella:2008}, while intercell fluxes are obtained by solving approximate Riemann problems with the Harten-Lax-van Leer-Einfeldt (HLLE) solver \citep{Harten:1982kq,Einfeldt:1988,Fields:2024pob}. A first-order flux correction is applied to ensure robustness in strongly shocked regions and to prevent unphysical states \citep{Lemaster:2009}. The induction equation is handled using constrained transport, which preserves the divergence-free condition $\nabla \cdot \mathbf{B} = 0$ to machine precision \citep{Evans:1988a,Gardiner:2007nc} in the absence of a stellar atmosphere (see Appendix \ref{app:conv}).
Time integration is performed using explicit Runge-Kutta schemes \citep{Gottlieb:2009a} with a Courant-Friedrichs-Lewy factor of $0.5$. After each timestep, primitive variables are recovered from the conserved variables via a nonlinear inversion procedure \citep{Kastaun:2020uxr}. Details of the primitive solver are provided in Appendix A of \citet{Cook:2023bag}.

\section{Results}
\label{sec:res}

\subsection{Magnetic field evolution}
\label{subsec:fields}

We begin by discussing the qualitative evolution of the magnetic field configuration.
As expected, the initial imposed magnetic field is subject to instabilities.  The Tayler instability manifests primarily in two modes: the ``varicose'' (or ``sausage'') mode and the ``kink'' mode \citep{Tayler:1957a,Lander:2011,Lasky:2012ju}. Each represents a characteristic pattern of deformation in magnetic flux tubes. 

\begin{figure}[t]
	\centering
	\includegraphics[width=0.49\textwidth]{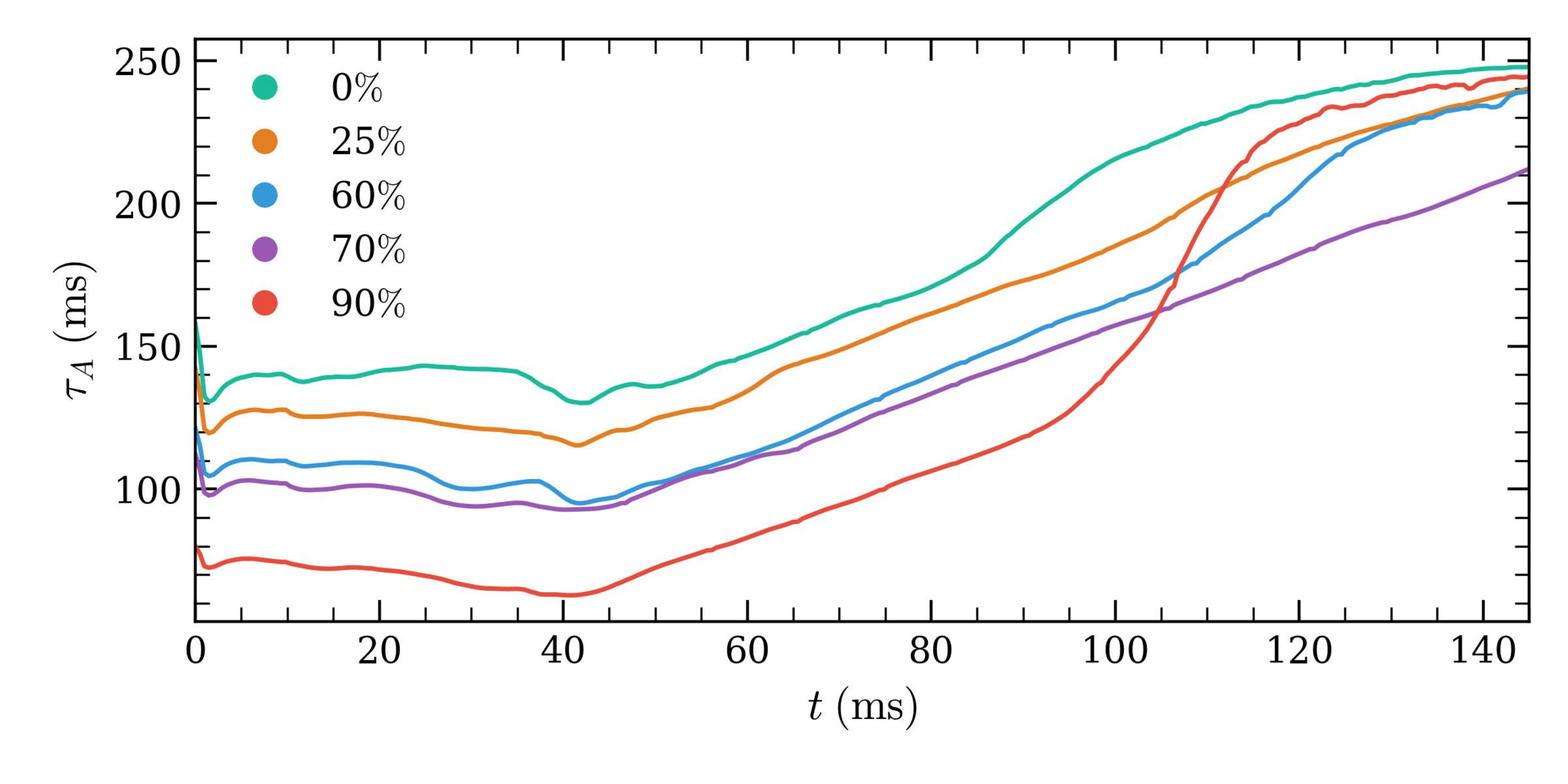}
	\caption{Alfv\'en time $\tau_A$ as defined in Eq. \eqref{eq:ta} versus the coordinate time for all GR models.}
	\label{fig:alfven1}
\end{figure}

In Figure~\ref{fig:2d} we show the evolution of the magnetic field magnitude $|\mathbf{B}|$ and poloidal field lines for Bt60.
As shown in the bottom left panel, the varicose instabilities arise in our simulations around $1\, \rm{ms}$ as radial contractions of the flux tubes. At this stage the effects of the sharp density gradient on the magnetic field lines are already significant. 
Kink instabilities, associated with the twisting of magnetic field lines, develop on the surface at around 
$10\, \rm{ms}$ following the transverse displacement of the fluid, before later emerging in the stellar interior. 

At $25\, \rm{ms}$, we see from the equatorial view that the field lines create vortex-like structures. 
The closed loops first emerge in the inner regions of the {$xz$} plane and subsequently migrate toward lower-density regions, as magnetic pressure drives them outward into the stellar envelope.
At about $50\, \rm{ms}$, the kink instability has saturated, and the highly asymmetric and unstable structure is dominated by small scale eddy features in both planes.

From this point onward our simulations show a nonlinear rearrangement of the field, where not only the inner closed loops are involved but the whole star and the field lines extending to the exterior.
At late times the field lines reach an axisymmetric dipole structure and the magnetic field strength is relatively homogeneous within the star.

The evolution of the magnetic field occurs on a characteristic timescale associated with the system, called the Alfv\'en crossing time. Since this quantity depends on the magnetic field strength $B$ which itself evolves in time, different definitions can be adopted. A discussion of these approaches is given in Appendix C of \cite{Cook:2025zzy}. In this work, we calculate the Alfv\'en time $\tau_A$ as proposed by \citet{Sur:2021awe} 
\begin{equation}
	\tau_A = \frac{2R \sqrt{4\pi \langle \rho \rangle}}{\langle B \rangle}
	\label{eq:ta}
\end{equation}
where $\langle \cdot \rangle$ represents volume-averaged quantities.
We show $\tau_A$ in Fig. \ref{fig:alfven1}, and we note that it varies across different cases due to the different initial magnetic energies. The initial values range between $50$ and $150\, \rm{ms}$ and it saturates to $220$ to $250\, \rm{ms}$ for all models on the simulated timescale as the magnetic field strength decays.

The Alfv\'en time is calculated in the stellar interior. We adopt a consistent definition of the stellar interior across all diagnostics. This definition is used both for the Alfv\'en time calculation and for all subsequent post-processing, including energetics and mode decomposition. Specifically, we use a fixed radius of $R_{int} \simeq 12.5\, \rm{km}$, since the initial TOV radius is rapidly altered by the strong magnetic fields. This choice of $R$ is implemented by defining the stellar interior as all regions with density above a prescribed threshold, $\rho\sim10^{12}\; \rm{g/cm^3}$ which yields the slightly larger effective stellar radius $R_{int}$. This same interior definition is used consistently throughout the analysis presented 

\begin{figure*}[t]
	\centering
	\includegraphics[width=1\textwidth]{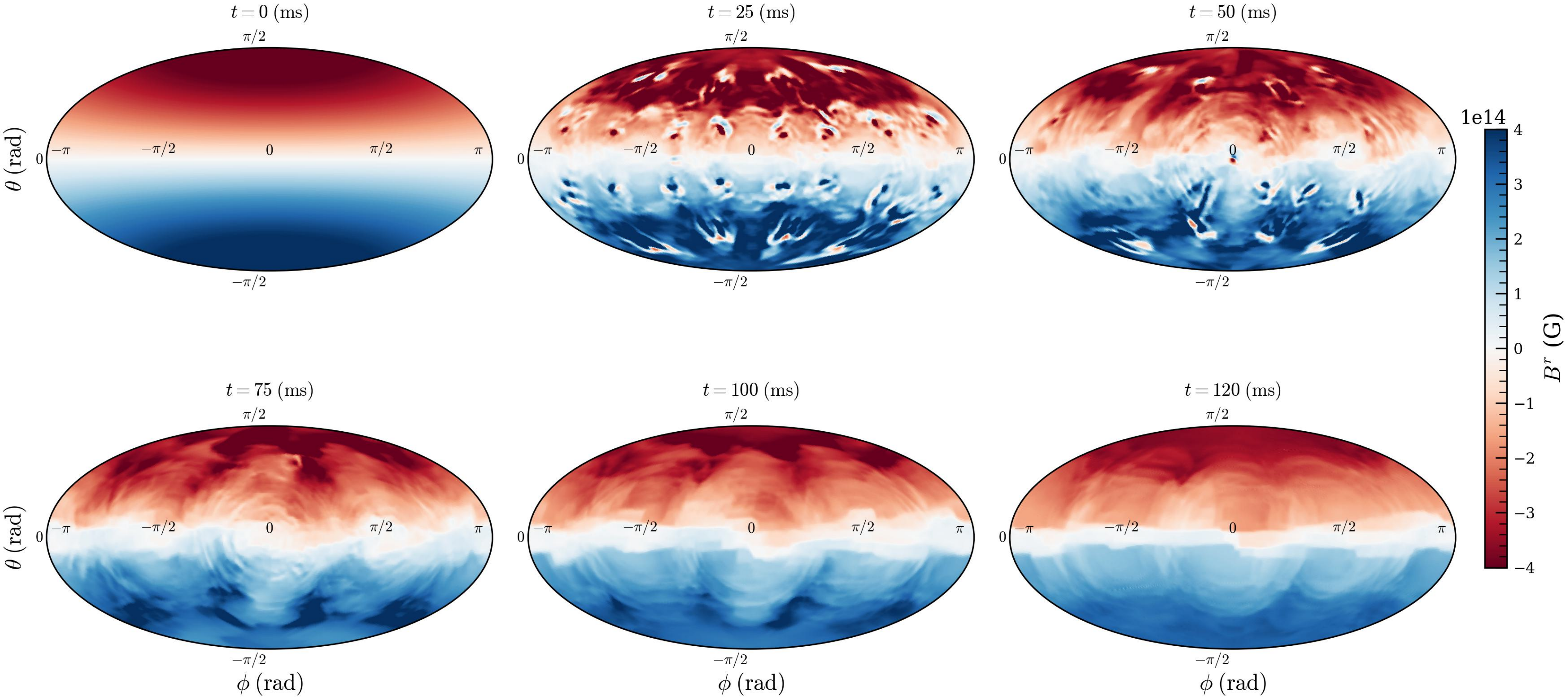}
	\caption{2D projections of the radial component $B^r$ on the stellar surface for Bt60. Times of the snapshots are given as figure titles.}
	\label{fig:surface}
\end{figure*}

In Figure \ref{fig:surface} we show the evolution of the radial magnetic field component $B^r$ plotted on the stellar surface.
The first snapshot corresponds to the initial condition, featuring outgoing field lines at the two poles, consistent with an aligned dipolar configuration. During the unstable phase driven by magnetic instabilities, small scale vortices develop at the surface leading to a distortion of the dipolar geometry. As the simulation evolves, the field partially reorganizes towards a dipolar configuration, while retaining non-axisymmetric features in the azimuthal ($\phi$) direction, particularly near the equatorial region.
We also note that the magnitude of $B^r$ increases during the unstable rearrangement phase before settling at late times to values comparable to the initially imposed field strength.

\subsection{Energetics}
\label{subsec:eg}
We now discuss the energy evolution of the models, focusing on the stability of the system and the role played by the magnetic field.
The decomposition of the energies performed is based on that discussed in Appendix B of \citet{Cook:2025zzy}.

Figure \ref{fig:energies} displays the energy evolution in the NS over the first $150\, \rm{ms}$ for all five models. We show the toroidal component of the magnetic field normalized by the total magnetic field strength and  the magnetic ($E_{mag}$) and kinetic ($E_{kin}$) energies normalized by the total energy.
The initial magnetic field energy depends on the initial toroidal field strength of the model $B_t$, as the surface magnetic field strength $B_p$ is equal in all simulations.
This setup is designed to explore the possible internal magnetic configuration, given a surface magnetic field $B_p$ inferred from dipole spin-down measurements.
\begin{figure*}[t]
	\centering
	\includegraphics[width=0.9\textwidth]{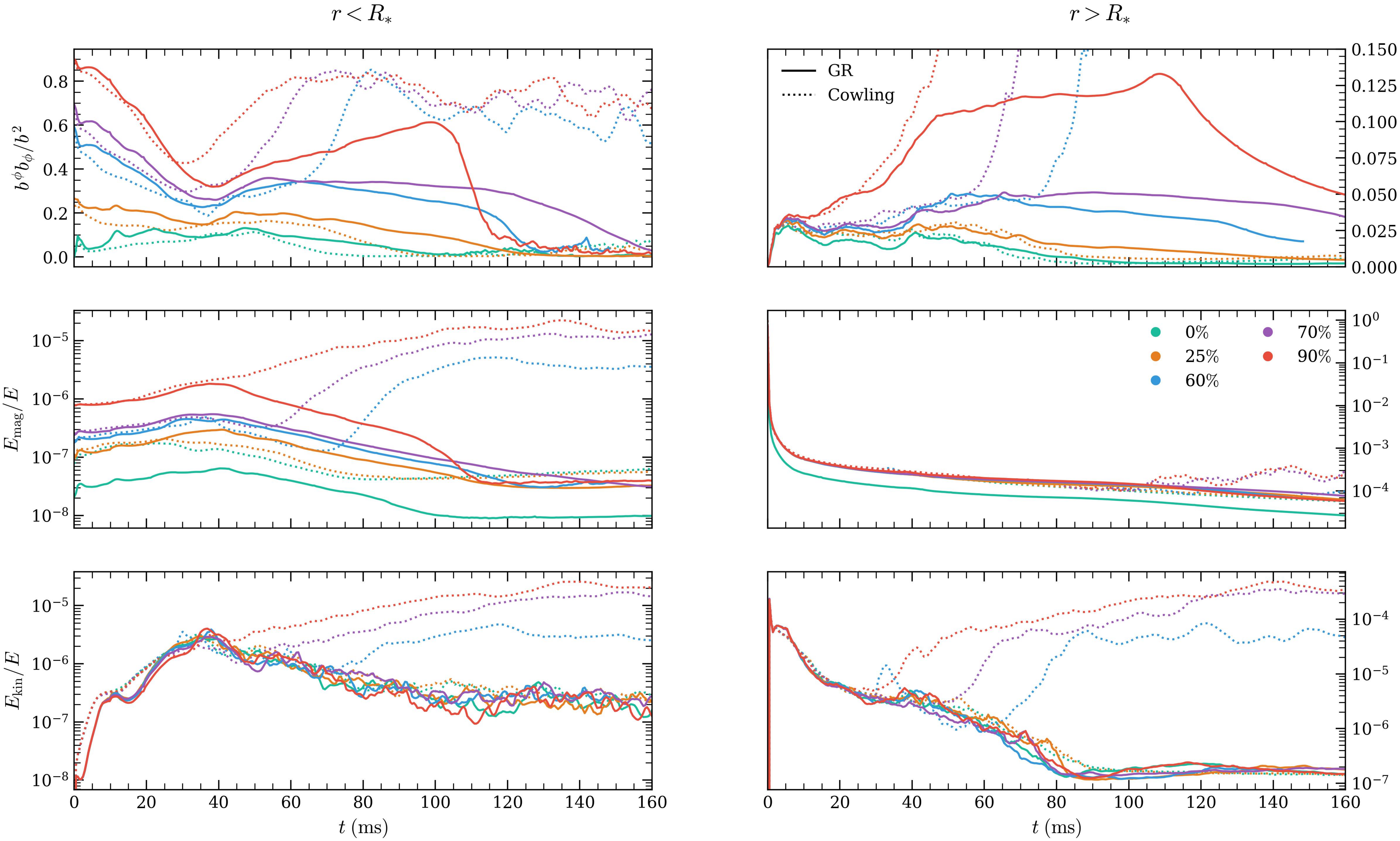}
	\caption{ Evolution of the energies for all configurations. (Upper) The evolution of the toroidal component of the magnetic field normalized by the total magnetic field strength. (Middle) Magnetic energy normalized by the total energy. (Lower) Kinetic energy normalized by the total energy. Left and right columns represent the energies in the interior and exterior respectively. Dotted lines represent the Cowling simulations.}
	\label{fig:energies}
\end{figure*}

A characteristic feature of our evolutions is the initial rearrangement of the magnetic field, due to our choice of initial conditions in which the field is strongest at the surface. The presence of a sharp density gradient across the stellar surface and high magnetization results in a sharp change in the toroidal magnetic energy within the first few milliseconds. Consequently, a leakage of the toroidal field into the exterior is observed, as shown in the upper-right panel of Fig. \ref{fig:energies}, due to the inability to accurately capture the sharp stellar surface in the presence of numerical diffusion, as commonly seen in previous studies \citep{Lasky:2011un}.

The kinetic energy rapidly grows in response to the initial readjustment of the magnetic field driven by the onset of the varicose instabilities described in Sec. \ref{subsec:fields}. In the first $40\, \rm{ms}$ the toroidal field of all models converges to values between $10 \%$ and $40 \%$. For the simulations with a stronger toroidal component, this results in a sharp decrease.
In the ideal MHD limit, the fluid is effectively tied to the magnetic field lines, so any evolution of the field forces the fluid to move with it. As a result, changes in the magnetic field can transfer energy into the motion of the fluid, increasing its kinetic energy.

As the kinetic energy reaches a first peak at $12\, \rm{ms}$, the varicose instabilities saturate and the kink instabilities begin to appear in the inner regions of the star.
During this rearrangement phase the magnetic energy does not decay immediately, but depending on the configuration, it increases by up to a factor of two.

While simulations of interior initially poloidal fields in non-rotating stars typically show an immediate decay of $E_{mag}$ independent of whether the initial field is poloidal, toroidal, or mixed \citep[e.g.][]{Sur:2021awe,Cook:2025zzy,Venturi:2025}, we observe a different behaviour during the rearrangement phase. In particular, the evolution at early times is qualitatively similar to the Newtonian results reported by \citet{Sur:2020hwn}, which employed the same initial conditions as those adopted here. A direct comparison is, however, limited by the shorter evolution times considered in their study. Like the models of \citet{Sur:2020hwn}, our initial configurations possess strong surface magnetic fields, which likely explain the initial increase in magnetic energy. These can alter how magnetic energy is redistributed during the early rearrangement phase and may delay the onset of global magnetic energy decay. 

By the time the overall kinetic energy peak is reached, around $40\, \rm{ms}$, the magnetic field has also peaked and shifted toward the center, with the kink instability dominating the dynamics. 
As closed field lines migrate toward the lower-density outer regions, the toroidal magnetic energy reaches the surface and begins to leak into the exterior. This results in the peak shown the upper-right panel in Fig. \ref{fig:energies}. At this stage, the system is highly dynamic, with both the kinetic and magnetic energy in the interior reaching their maximum values for all GR models.

Subsequently, the energy evolutions of models evolved in full GR and in the Cowling approximation begin to differ. The Cowling models with a strong toroidal component, ranging from $60\%$ to $90\%$, succumb to instabilities and transition to a turbulent and highly energetic state. The surface instabilities grow exponentially, triggering a dynamical response that enhances the internal toroidal field and kinetic energy. This results in the NS expansion to up to $13.5\, \rm{km}$, causing part of the stellar material to enter the region defined as the exterior and thereby contributing to the observed increase in the exterior magnetic and kinetic energies. After $80\,\rm{ms}$, this leads to mass loss through the domain boundaries resulting in a poorer mass conservation of 1 part in $10^6$.
In the Cowling approximation the metric remains fixed and spherical, which is not compatible with the evolving magnetic field geometry. As the magnetic energy evolves, the energy associated with the gravitational field cannot correspondingly adjust. As a consequence, excess magnetic energy can be artificially retained or redirected within the fluid evolution, enhancing the growth of instabilities. In contrast, gravity feedback naturally couples the spacetime to the matter and magnetic fields, which acts to suppress the development of these instabilities. In the Cowling models with a smaller toroidal component the magnetic fields are not strong enough to trigger this response. A similar behavior was observed in \cite{Sur:2021awe}, where a Cowling simulation with an initial toroidal field of $80\%$ of the total magnetic energy showed the toroidal component first decreasing and then recovering its initial value. By contrast, \citet{Sur:2021awe} simulations initialized with a purely poloidal field developed a toroidal component that subsequently decayed to approximately $1\%$.

Focusing on the GR models, after $40\, \rm{ms}$ the simulations enter a nonlinear relaxation phase in which both magnetic and kinetic energies decrease steadily. This behavior does not appear to be primarily driven by the development of instabilities, as seen in \citet{Cook:2025zzy}, but is instead consistent with the leakage of toroidal magnetic flux into the atmosphere. This effect becomes particularly significant at $t \sim 40\,\rm{ms}$, coinciding with the transition point in both $E_{\rm mag}$ and $E_{\rm kin}$. At the same time, we observe a small peak in kinetic energy in the exterior, as shown in the lower-right panel.
In this highly nonlinear phase, the maximum internal toroidal field reached by the models ranges between $10\%$ and $35\%$ of the total magnetic energy. Only Bt90 shows sustained amplification, developing a maximum toroidal fraction of $60\%$. In contrast to the Cowling evolutions, at around $100,\rm{ms}$ this model begins to decay and eventually settles to values similar to the other runs. This behavior indicates that, especially in the presence of strong magnetic fields, back reaction of the fully dynamical gravitational field plays a non-negligible role in regulating the dynamics.
The internal magnetic energy in the GR evolutions drops from its initial value to between ${\sim}3\%$ to ${\sim}50\%$ depending on the initial configuration, after which it remains approximately steady for the rest of the evolution. The evolution of $E_{\rm mag}$ over the full domain closely follows that of the internal component, since the contribution from the exterior remains negligible in comparison.

The decay of magnetic energy in both the interior and exterior regions is primarily associated with the physical rearrangement of the magnetic field and the expulsion of toroidal flux, rather than being purely a numerical artifact. Nevertheless, finite-resolution effects likely contribute to the long-term secular dissipation observed at late times.
As explained in \citet{Cook:2025zzy}, we do not expect a perfect conservation of internal energy, and we obtain $\Delta E_{int} \approx 10^{-4}$ as shown in Appendix \ref{app:conv}.
This is the result of the combination of an improved atmosphere treatment and conserved-to-primitive approach, detailed in the appendix of \citet{Cook:2025zzy}, and the first order flux correction implemented in \AK, detailed in \citet{Fields:2024pob}. 

As the field approaches a quasi-stationary global scale configuration by $150\, \rm{ms}$, the toroidal magnetic energy has reached a magnitude between $0.5 \%$ and $10\%$ for all models. Our findings agree with the mixed configurations reported in \citet{Sur:2020hwn} over early timescales, although our simulations extend for a longer duration. As their evolutions maintain a fixed gravitational field, we would expect the toroidal field to grow again on longer timescales by a similar energy balance argument to that above.
Similarly, \citet{Cook:2025zzy} reported that purely poloidal, interior-confined configurations generate a toroidal contribution reaching values of ${\sim}15\%$ over similar timescales. These values are comparable to the maximum reached by our purely poloidal configuration. However, in their setup the toroidal component remained stronger, since the magnetic field was constrained within the interior and did not leak into the atmosphere, which limited its decay. 
Our results are also consistent with the non-rotating model presented in \citet{Venturi:2025}. As discussed above, our Cowling runs are consistent with the long-term evolution of the purely poloidal setup studied in \citet{Sur:2021awe}, and show qualitatively similar behaviour to their strongly toroidal configuration, with discrepancies arising from the different magnetic field configurations adopted here, where stronger fields extend across the stellar surface.

\begin{figure}[t]
	\centering
	\includegraphics[width=0.49\textwidth]{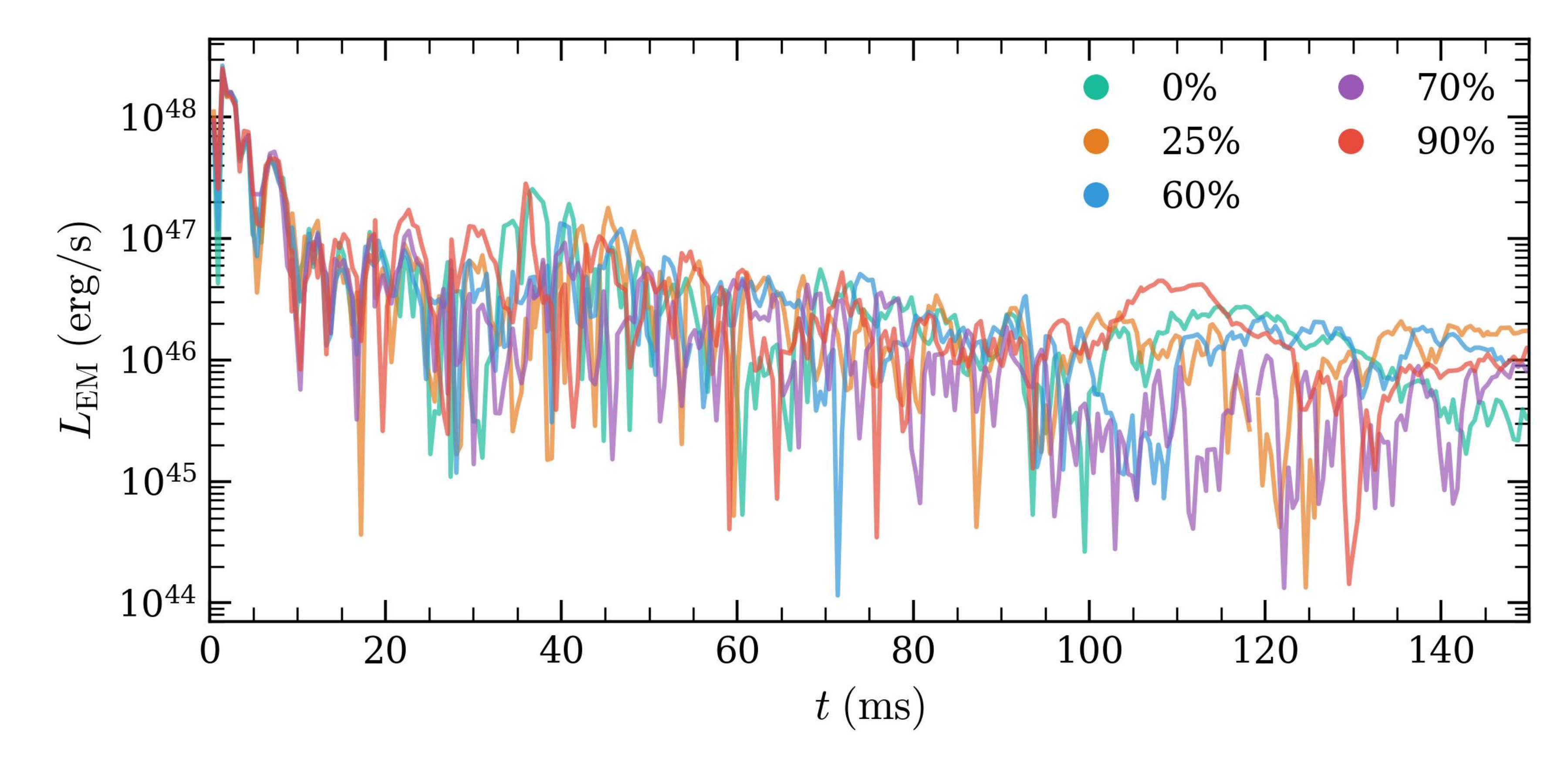}
	\caption{Time evolution of the electromagnetic luminosity $L_{\rm EM}$ computed on a spherical shell at $R_{int}$, for all models considered.}
	\label{fig:lum}
\end{figure}

The electromagnetic luminosity $L_{\rm EM}$, computed from the Poynting flux through a spherical shell at the stellar surface, reflects the different stages of the magnetic field evolution. As shown in Fig. \ref{fig:lum}, all models exhibit an initial burst during the first few milliseconds associated with the rapid magnetic field readjustment near the stellar surface. Following this transient, the luminosity decreases by roughly one order of magnitude as the interior field reorganizes. When the toroidal magnetic field leaks from the surface at around $40\, \rm{ms}$, the models present a small increase in the electromagnetic emission. At later times the luminosity settles into a broader quasi-stationary phase with values of $L_{\rm EM} \sim 10^{46}\; \rm{ erg/s}$, concurrent with the development of a more stable large scale magnetic configuration.

\subsection{Mode analysis}
\label{subsec:modes}
We measure the radially integrated angular distribution of two quantities, the density $\rho$ and the radial magnetic field component $B^r$. We analyze a full decomposition into spherical harmonics differentiating between the interior and exterior.

We express for a quantity of interest $f$, the decomposition
\begin{eqnarray}
	a_{\ell,m}(f) &=& \int f Y^*_{\ell,m}(\theta, \phi) r^2 dr d\Omega \label{eq:alm}
\end{eqnarray}
\begin{figure}[b]
	\centering
	\includegraphics[width=0.49\textwidth]{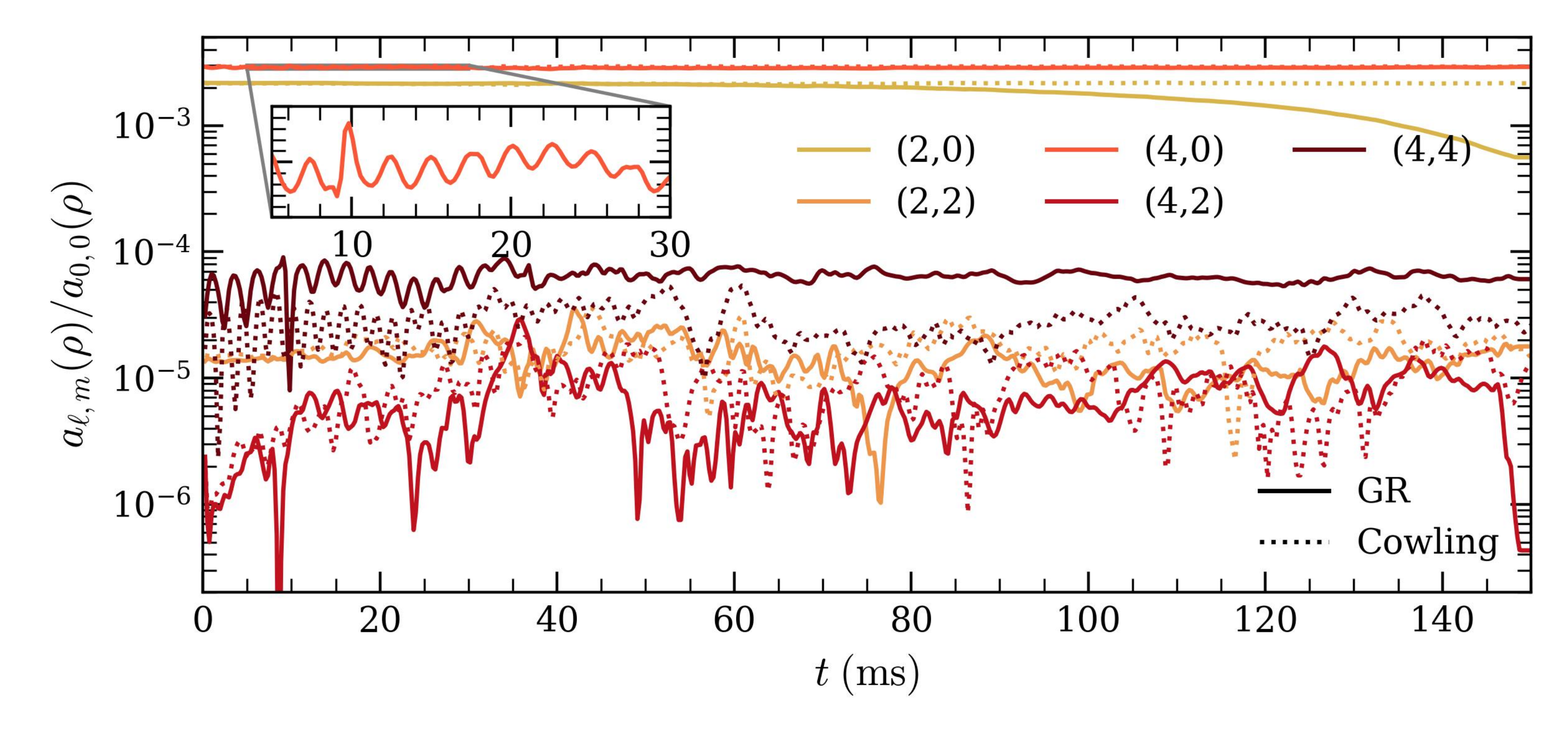}
	\caption{Evolution of the spherical harmonic mode contributions to density profile in the stellar interior for Bt60 in both GR and Cowling. Modes are normalized by the coefficient $a_{00}$, which is the overall dominant mode. Inset: Zoom in on the $(4,0)$ mode.}
	\label{fig:almrhoint}
\end{figure}

In Figure \ref{fig:almrhoint}, we show the spherical harmonic mode contributions to the density profile in the interior for a single representative model, since all simulations exhibit a comparable evolution. For $\rho$ we include only dominant modes $(\ell,m) = (2,0), (2,2), (4,0), (4,2), (4,4)$. 
\begin{figure}[t]
	\centering
	\includegraphics[width=0.49\textwidth]{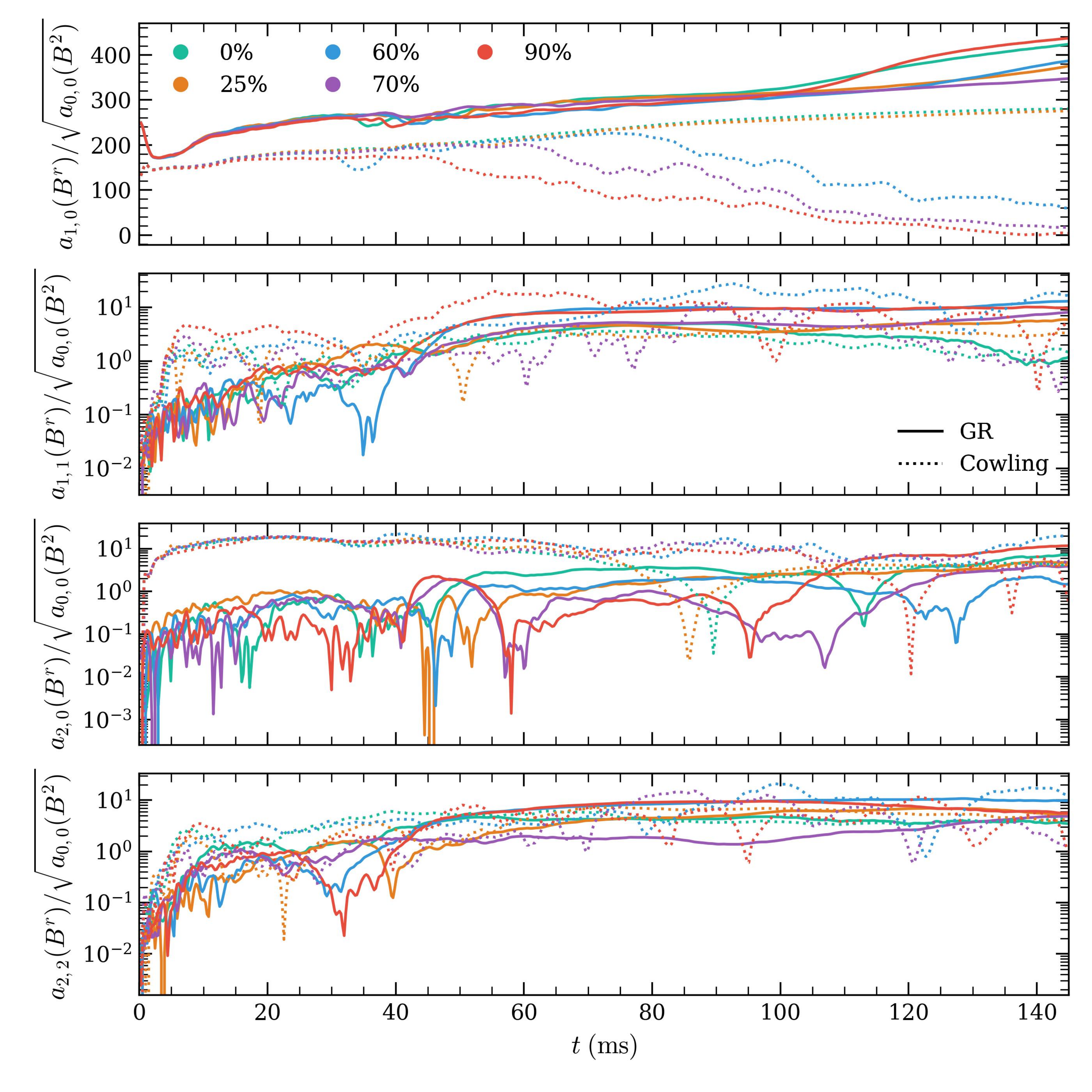}
	\caption{Evolution of the spherical harmonic mode contributions to the radial magnetic field strength $B^r$, calculated in the exterior. Modes are normalized by the square root of the coefficient $a_{0,0}$ of the total magnetic field strength $B^2$.}
	\label{fig:almbr}
\end{figure}

Of the non-axisymmetric modes, the $(4,4)$ mode mostly dominates in magnitude, consistent with the Cartesian grid seeding of the instability. The mode exhibits oscillatory behavior during the rearrangement phase before stabilizing, while remaining dominant over the other non-axisymmetric components. As discussed in Sec. \ref{subsec:fields}, small scale features emerge on the surface during the rearrangement phase (see Fig. \ref{fig:surface}). 
We associate these features with excitations of the $(4,4)$ mode, and note that, after $t=50\; \rm{ms}$, when the surface features have dissipated, that the oscillatory behaviour in this mode is also suppressed.
The $(2,2)$ mode is the second most dominant and remains constant, while the $(4,2)$ mode grows significantly throughout the evolution. The $m=2$ sector exhibits stronger growth in the Cowling simulations and, at late times, reaches amplitudes comparable to the $(4,4)$ mode.

Focusing on the axisymmetric sector, both $(2,0)$ and $(4,0)$ modes remain dominant throughout the evolution. The inset in Fig. \ref{fig:almrhoint} shows that the $(4,0)$ mode also exhibits small scale oscillations during the rearrangement phase, similar to the $(4,4)$ mode discussed above. Only in the GR case does the $(2,0)$ mode show a late-time decay, which is attributed to GW emission, as discussed in Sec. \ref{subsec:gw}.
The oscillatory signals observed in the $(4,4)$ and $(4,0)$ modes are also present in the non-magnetized simulation, indicating that the density oscillations drive a corresponding response in the magnetic field, rather than being induced by it.

In Figure \ref{fig:almbr} we show the spherical harmonic mode contributions to the radial magnetic field strength $B^r$ calculated in the exterior. The initial field rearrangement disrupts the poloidal field, affecting the $(1,0)$ mode greatly. This then steadily increases throughout the evolution, in agreement with the rearrangement of the field lines toward an axisymmetric configuration. Around $40\, \rm{ms}$ this mode shows some detachment from the linear growth in response to the toroidal field leak from the surface discussed in the previous section. 

The strongly toroidal Cowling runs behave differently, as they do not evolve towards an axisymmetric dipolar structure.
Since in the Cowling approximation the spacetime metric is fixed to remain spherically symmetric, the system is driven towards spherical symmetry, thereby suppressing the $\ell=1$ component.  This behavior is most prominent for the cases with strong magnetic energies that violate the symmetry of the metric.

In the GR case, the $(1,1)$, $(2,0)$ and $(2,2)$ modes grow exponentially at early times and then stabilize with similar magnitudes. In the Cowling case, the $(1,1)$ and $(2,0)$ modes grow much faster at early times but reach the same final magnitude.

\begin{figure}[t]
	\centering
	\includegraphics[width=0.49\textwidth]{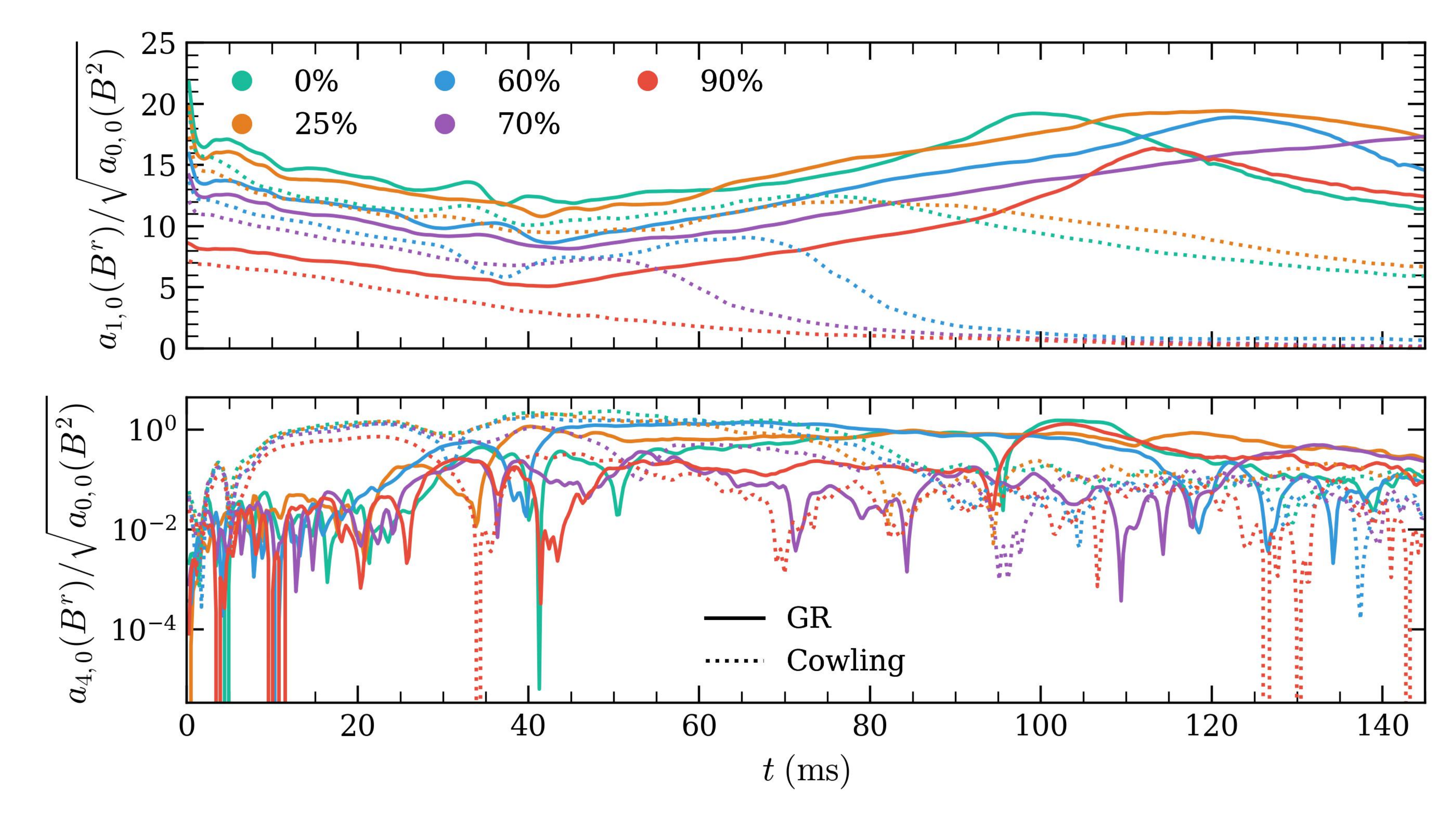}
	\caption{Evolution of the spherical harmonic mode contributions to the radial magnetic field strength $B^r$, calculated in the interior. Modes are normalized by the square root of the coefficient $a_{0,0}$ of the total magnetic field strength $B^2$.}
	\label{fig:almbr2}
\end{figure}

To further analyze the features seen in Fig. \ref{fig:surface}, we show the interior spherical harmonic mode contributions to $B^r$ for $(1,0)$ and $(4,0)$ in Fig. \ref{fig:almbr2}. During the time of magnetic field rearrangement, the $(1,0)$ mode exhibits a phase of decay as the dipole field is disrupted by the instabilities, while the $\ell = 4$ modes show a corresponding growth, consistent with the emergence of surface structures. After the transition phase at $t = 40\, \rm{ms}$, the $(1,0)$ mode begins to grow again leading to the gradual dissipation of these features. These results are consistent with what is seen in the projections of $B^r$. Here we again observe the decay of the $(1,0)$ mode for the unstable configurations.

\begin{figure}[b]
	\centering
	\includegraphics[width=0.49\textwidth]{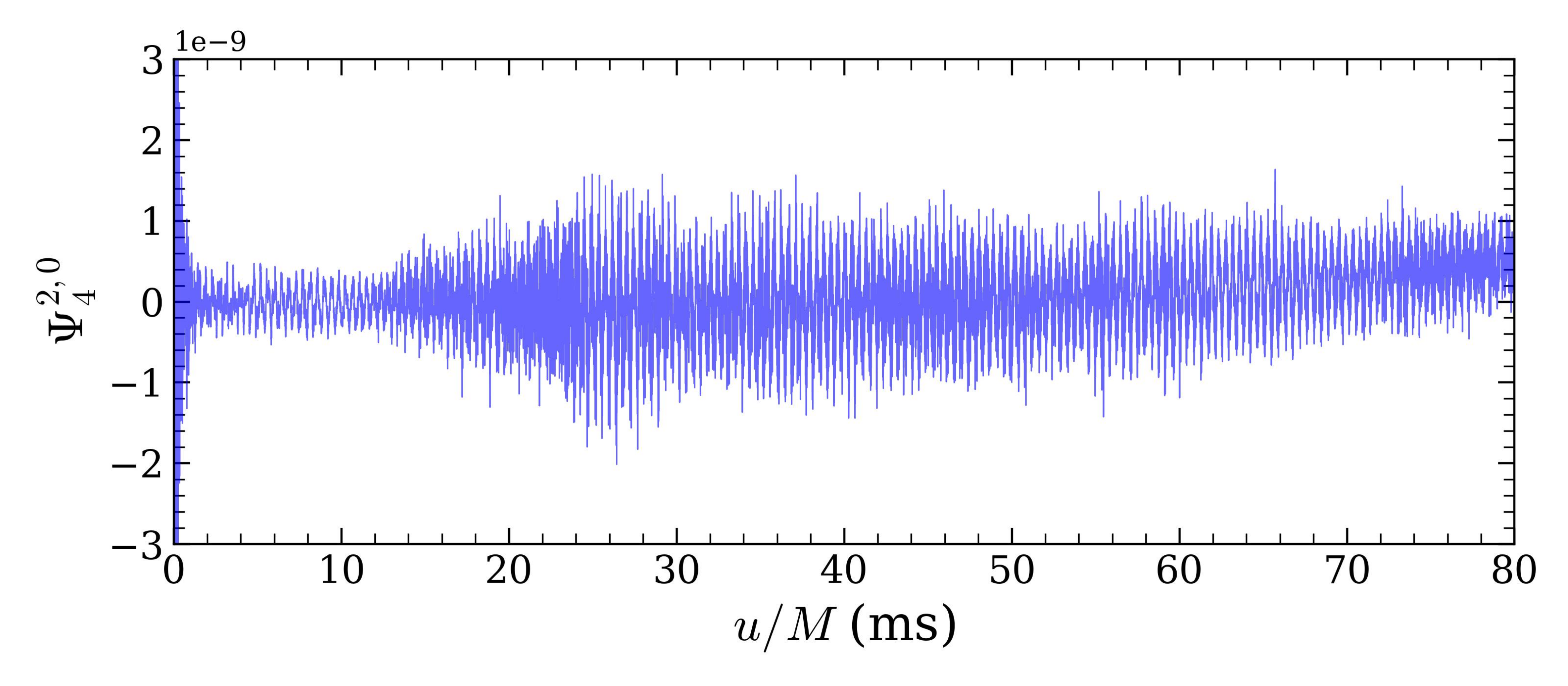}
	\caption{GW signal of Bt60 extracted at $R_{gw} = 150\, \rm{km}$ as a function of normalized retarded time.}
	\label{fig:psi4}
\end{figure}

\subsection{Gravitational wave analysis}
\label{subsec:gw}

In \AK, GWs are extracted by computing the Newman-Penrose Weyl scalar $\Psi_4$ at fixed coordinate radii.
The GW signal is then decomposed into spin-weighted spherical harmonic modes, with particular focus on the dominant $(\ell, m) = (2,0)$ and $(2,2)$ components. We expect the $(2,0)$ and the $(2,2)$ modes to be the strongest contributors to the GWs. The former captures the dominant axisymmetric oscillations induced by the magnetic-field evolution, while the latter probes non-axisymmetric dynamics associated with the instability development.

In Figure \ref{fig:psi4} we show the $(2,0)$ gravitational waveform extracted at a radius of $R_{gw}  = 150\, \rm{km}$ plotted as a function of the retarded time $u/M=(t-R_{gw})/M$. This choice of $R_{gw}$ balances extraction in the wave zone with resolution and boundary effects that can contaminate the signal and cause drift at larger radii \cite{Fontbute:2025ixd}.

The waveform exhibits distinct phases that reflect the underlying dynamics within the star: stronger and more rapidly evolving deformations produce more intense GW emission. During the initial phase, the instability of the magnetic field leads to a growth in the amplitude of $\Psi_4$ reaching a peak at approximately $25\, \rm{ms}$, in agreement with the growth observed in both the magnetic and kinetic energies. The subsequent decay of these energies is also reflected in the signal, which develops a low frequency beating pattern. However, resolving this modulation requires a longer lasting signal.

\begin{figure}[t]
	\centering
	\includegraphics[width=0.49\textwidth]{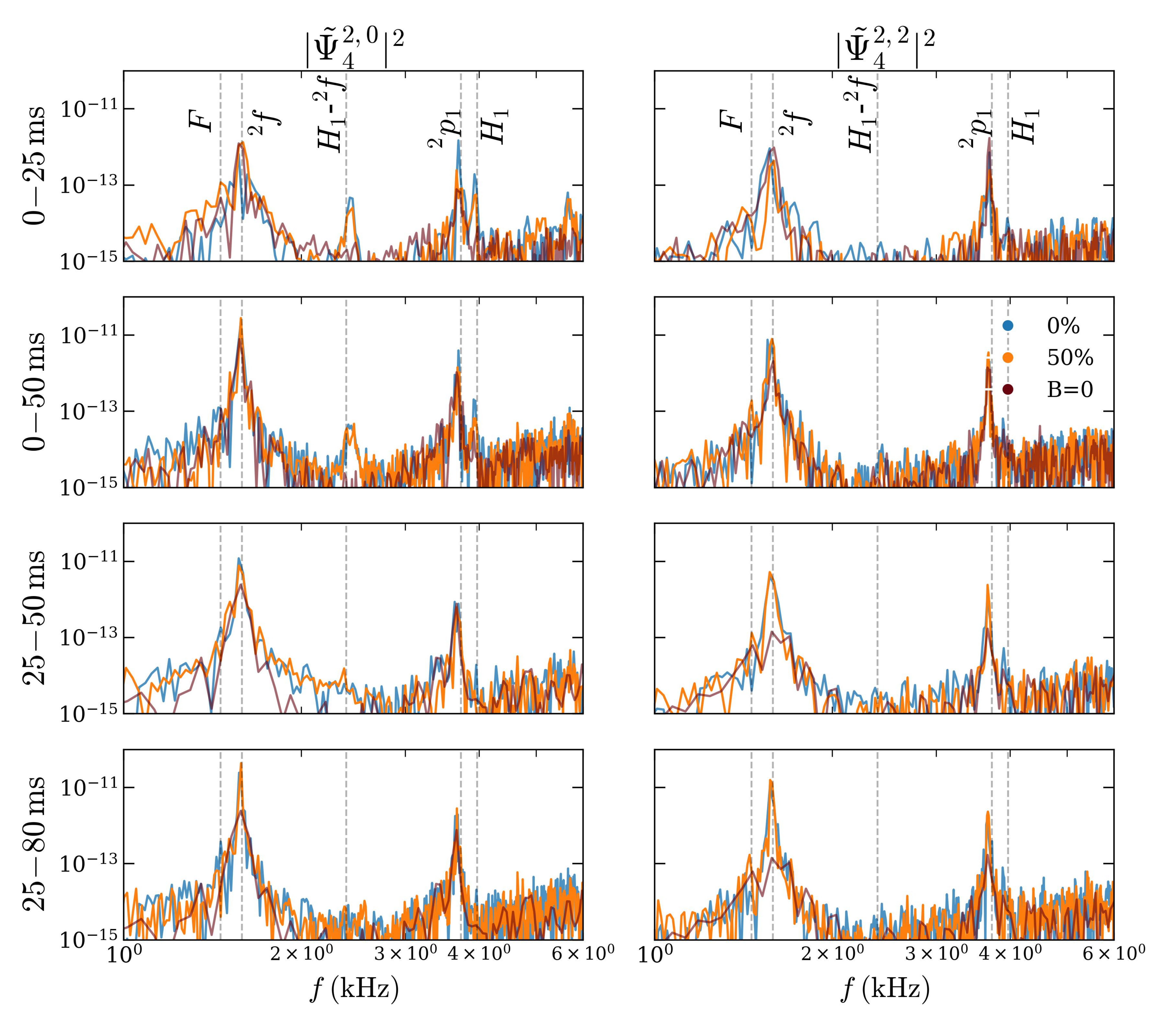}
	\caption{Power spectral density (PSD) of $\Psi_4^{2,0}$ (left panels) and
          $\Psi_4^{2,2}$ (right panels) for B0, Bt0, Bt60. Each row of two panels refers to the PSD over selected time windows indicated on the ordinate. The PSD is in arbitrary units.}
	\label{fig:phase}
\end{figure}

Fourier analysis is applied to the evolutionary phases of $\Psi_4$ to identify characteristic frequencies and mode mixing, which are indicative of the system’s oscillatory behavior.
The analysis is carried out separately for the multipole components $(\ell,m)=(2,0)$ and $(\ell,m)=(2,2)$, denoted respectively as $\Psi_4^{2,0}$ and $\Psi_4^{2,2}$. The results are shown in Fig. \ref{fig:phase}, where we compare the peak frequencies in the spectrum of Bt0 and Bt60 to the non-magnetized case B0 to quantify the influence of magnetic fields on the star's dynamics.

We observe that the main spectral peaks are consistent with the expected eigenmode frequencies of non-magnetic NS \citep{Dimmelmeier:2005zk,Baiotti:2008nf}, where $F$ is the frequency of the fundamental quasi-radial $(\ell = 0)$ mode, ${H}_1$ its first overtone, ${}^2\!f$ the fundamental pressure quadrupole $(\ell=2)$ mode and ${}^2p_1$ its first overtone.

However, the magnetized runs present an additional peak at early times ($0-25\, \rm{ms}$) that does not appear in the non-magnetized case. This feature is compatible with the nonlinear coupling ${H}_1$-${}^2\!f$ between the $\ell=2$ fundamental pressure mode ${}^2\!f$ and the first quasiradial overtone ${H}_1$ \citep{Baiotti:2008nf} excited at early stages of the dynamics. We note that this peak is more prominent for Bt60 than Bt0, suggesting that the coupling is triggered by the toroidal field evolution. The ${}^2\!f$ channel has the strongest power in the spectrum while ${H}_1$ is the most significant channel for quasi-radial oscillations with a larger peak than the ${F}$ mode (see upper left panel of Fig. \ref{fig:phase}).
This $\ell=0$ and $\ell=2$ coupling is compatible with the $a_{\ell m}$ projections shown in Fig. \ref{fig:almrhoint}.

Overall, these results suggests that the magnetic instabilities may imprint the GW spectrum with specific nonlinear mode couplings. Similar effects may be expected in GWs from newly born proto-NSs after core-collapse supernovae, though the latter might be challenging to identify even with third-generation detectors \citep{Ott:2008wt,Afle:2023}.

\section{Discussion and conclusions}
\label{sec:con}

In this work, we presented long-term GRMHD simulations of NSs with an external dipole field and a mixed poloidal-toroidal internal field. To our knowledge, ours is the first work considering such magnetic field configurations in general relativity on such timescales. We investigated the nonlinear evolution of initially unstable magnetic field configurations and their relaxation toward quasi-equilibrium states under fully dynamical spacetime evolution, as well as under the Cowling approximation for comparison.

We have found that initially unstable magnetic fields evolve through Tayler instability-driven nonlinear dynamics, leading to a robust relaxation toward an approximately stable mixed poloidal-toroidal equilibrium. Across all cases studied, the system converges toward a predominantly axisymmetric, dipole-like configuration with structures in the azimuthal direction. The final magnetic structure retains a small but persistent toroidal component in the range of $0.5\%$ to $10\%$ of the total magnetic energy, concentrated in the equatorial region.
Despite differences in the initial magnetic field topology, the systems converge to similar quasi-stationary end states, suggesting that the final equilibrium is largely insensitive to the initial conditions within the class of configurations explored. This behavior supports the idea that strongly magnetized NSs may evolve toward a limited set of equilibria under ideal GRMHD evolution. 

A key role is played by the exchange of magnetic energy across the stellar surface, which significantly influences the dynamics of the internal magnetic field rearrangement. Whether the toroidal field remains confined or partially leaks into the atmosphere has direct implications for the long-term magnetic structure and stability of the star. The resulting configurations also leave observable imprints: both electromagnetic emission and GW signals carry signatures of the magnetic instability-driven dynamics. We measure a late time electromagnetic luminosity settling at approximately $L_{\rm EM} \sim 10^{46} \; \rm{ erg/s}$, broadly consistent with expectations for strongly magnetized NS systems.

The GW signals considered here correspond to long-lived, approximately monochromatic emission from NS oscillation modes.
However, rapid magnetic-field rearrangements or fluid readjustments can introduce temporary spectral evolution, producing transient features. These are relevant to highly dynamical scenarios such as magnetar flares, glitches, or post-merger remnants.
While continuous-wave emission from secular magnetic deformations is beyond the scope of this work, the magnetic field evolution studied here is also relevant for determining the long-term GW-emitting quadrupolar structure of NSs (for a review see, e.g., \citet{Lasky:2015rev}).

Our results are consistent with the Newtonian simulations of \citet{Sur:2020hwn}, although here we employ a significantly larger computational domain and general-relativistic evolution, allowing us to reduce boundary effects and evolve the system over longer timescales. These improvements reveal a more complete picture of the late time relaxation of the magnetic field. In particular, the larger grid size reduces spurious boundary feedback and allows partial escape of toroidal magnetic flux into the external medium. Comparing with \citet{Sur:2021awe} and \citet{Cook:2025zzy}, our results indicate that the development and saturation of the toroidal magnetic component are strongly sensitive to whether the field is concentrated in the interior or extends outside the stellar surface. In particular, configurations that permit exchange with an external region lead to weaker toroidal fields.

The comparison between Cowling and fully dynamical GRMHD simulations highlights the role of spacetime dynamics in the redistribution of magnetic energy and in the timescale of relaxation toward equilibrium. While the Cowling approximation captures the qualitative development of the early instability, fully dynamical simulations provide a more complete description of energy redistribution and long-term settling behavior.

In particular, in the Cowling approximation, models with a strong toroidal component develop pronounced surface instabilities that lead to a strong growth of the magnetic field, enhanced magnetic energy redistribution, and magnetic flux leakage into the exterior. This is due to the inability of the spacetime to adapt to the geometry of the strong magnetic energy density. This can ultimately result in degraded mass conservation due to boundary outflows at late times. In contrast, fully dynamical GR simulations, which allow back reaction of the matter content onto the spacetime geometry, show a smoother nonlinear relaxation toward quasi-equilibrium states with a small toroidal field.

Future work will extend this study to twisted-torus configurations and rotating NSs. The latter will allow us to determine how rotation influences the generation of toroidal magnetic fields. Rotation has been shown to play a crucial role in both the stability and geometry of the magnetic field and enables a more direct connection to observed NS populations \cite{Tsokaros:2021pkh}. In particular, including rotation leads to more prominent GW emission, potentially further impacting the rearrangement of the magnetic field.

A limitation of this work is the absence of resistive MHD effects. In the ideal MHD framework, reconnection is not physically modeled but only arises through numerical dissipation. This may affect the efficiency of magnetic topology rearrangement and could influence the quantitative fraction of toroidal field surviving at late times, see \citet{Cheong:2024stz} for recent results in this direction. Future improvements should therefore include resistive or extended MHD formulations to better capture reconnection and magnetic dissipation processes.

Additionally, a more realistic treatment of the stellar atmosphere and exterior plasma would help clarify the degree of magnetic flux leakage and its observational consequences. We use an ideal gas EOS without explicitly controlling the stellar matter stratification, which may affect the stability of the magnetic configurations as discussed by \citet{Becerra:2022}. Longer simulations would also be valuable to confirm whether the observed quasi-stationary states are truly asymptotic equilibria or long-lived transient states. On a technical level, such simulations remain challenging due to the numerical treatment of the magnetized stellar atmosphere.

\section*{acknowledgments}

AC, SB, BH acknowledge support for the MERLIN project. MERLIN is funded by the Deutsche Forschungsgemeinschaft (DFG; BE 6301/6-1 Projektnummer: 524726453) and the Narodowe Centrum Nauki (NCN) OPUS-LAP grant number 2022/47/I/ST9/01494, under the EU weave initiative. SB acknowledges funding from the EU Horizon under ERC Consolidator Grant, no. InspiReM-101043372. JF acknowledges support from NASA under award No. 80NSSC25K7213.We acknowledge the EuroHPC Joint Undertaking for awarding this project access to the EuroHPC supercomputers, LEONARDO, hosted by CINECA (Italy) and the LEONARDO consortium; LUMI, hosted by CSC (Finland) and the LUMI consortium; and KAROLINA hosted by IT4Innovations National Supercomputing Center (Czech Republic) through an EuroHPC Benchmark Access call (EHPC-BEN-2024B10-018). 
Computations were also performed on the ARA and DRACO clusters at Friedrich Schiller University Jena. The ARA cluster is funded in part by DFG grants INST 275/334-1 FUGG and INST 275/363-1 FUGG, and ERC Starting Grant, grant agreement no. BinGraSp-714626. We acknowledge ISCRA for awarding this project access to the LEONARDO supercomputer, owned by the EuroHPC Joint Undertaking, hosted by CINECA (Italy) through projects IsCc8\_MERLIN0 and IsB31\_MERLIN1. The authors gratefully acknowledge the Gauss Centre for Supercomputing e.V. (www.gauss-centre.eu) for funding this project by providing computing time on the GCS Supercomputer JUWELS \citep{JUWELS:2021a} at Jülich Supercomputing Centre (JSC) through project MAGWAVE. The authors gratefully acknowledge the Gauss Centre for Supercomputing e.V. (www.gauss-centre.eu) for funding this project by providing computing time on the GCS Supercomputer SuperMUC-NG Phase 2 at Leibniz Supercomputing Centre (www.lrz.de) through projects pn67xo and pn76li.
AC thanks Raj Kishor Joshi and Jorge L. Rodr\'{i}guez-Monteverde for helpful comments and discussions. AC is grateful to S. and M. Capobianco for continuous support during the development of this work.

\section{Appendix: Resolution and boundary effects}
	
\label{app:conv}

\begin{figure}
	\centering
	\includegraphics[width=0.49\textwidth]{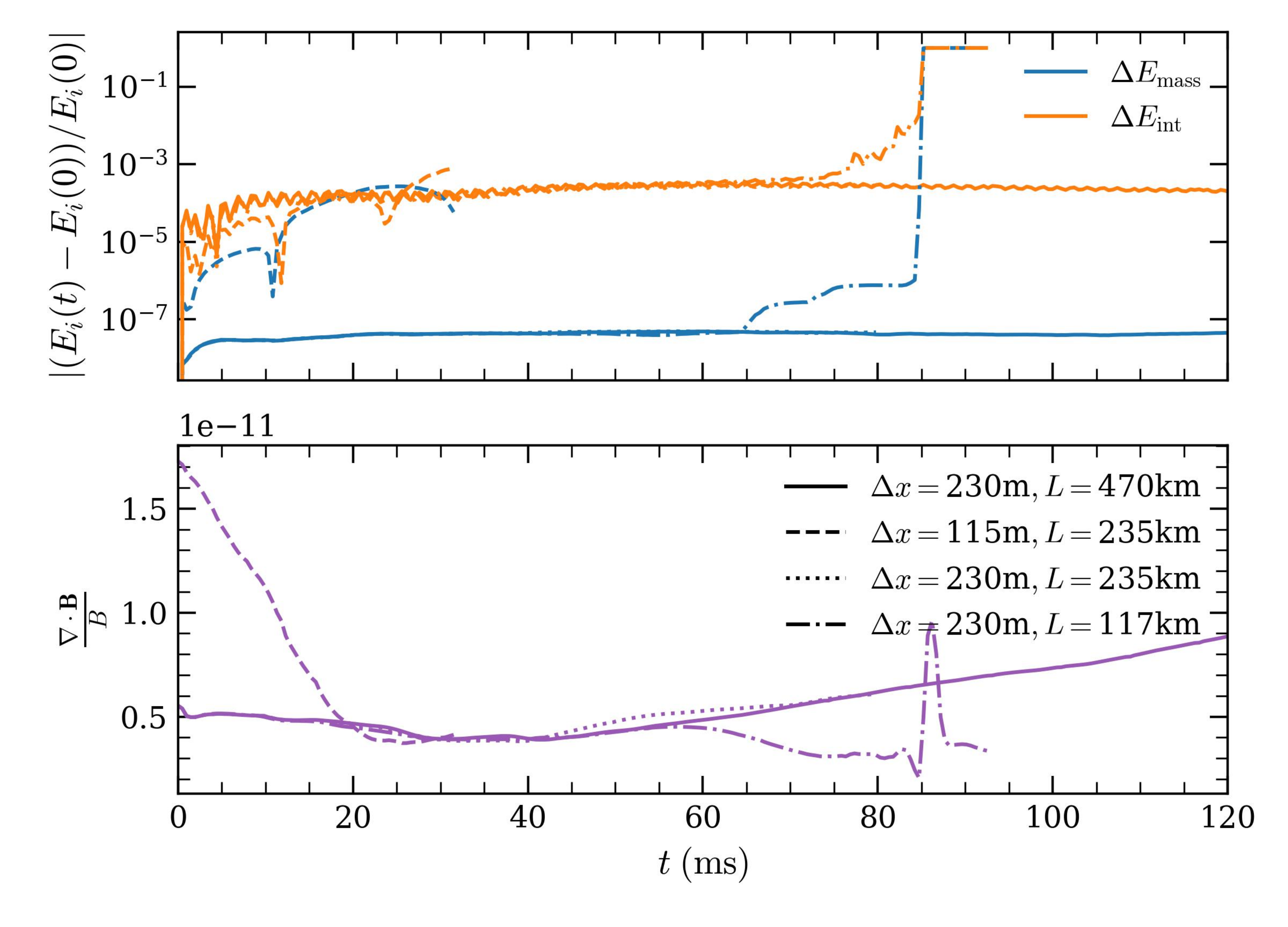}
	\caption{(Upper) Relative variation in each energy component, $\Delta E_i$, normalized by its initial value $E_i(t=0)$, for the rest-mass, internal, and kinetic energies. (Lower) Divergence of the magnetic field. Different line styles correspond to different outer boundary distances and resolutions.}
	\label{fig:diag_conv}
\end{figure}

We perform a series of simulations varying the distance between the outer boundary and the origin, $L$, to investigate the impact of boundary effects on our simulations. We perform these for $L = [117,235,470]\, \rm{km}$, with the resolution on the star fixed to $\Delta x = 230\, \rm{m}$, in addition to a higher resolution simulation with $\Delta x = 115\, \rm{m}$ and $L = 235\, \rm{km}$ . In Fig. \ref{fig:diag_conv} we show the evolution of $\Delta E_{int}$, $\Delta E_{mass}$ and $\nabla \cdot \mathbf{B} / B$ for all setups.

Our simulations feature strong magnetic fields extending into the atmosphere, thus influencing the entire computational domain. If the outer boundary is placed too close to the NS, the imposed boundary conditions can contaminate the evolution, as observed in the $L = 117\, \rm{km}$ case. To prevent spurious boundary effects from propagating into the spacetime evolution we adopt a larger domain size of $L = 470\, \rm{km}$ in all production runs.

As described in \citet{Fields:2024pob}, in Eulerian simulations, many numerical steps require dividing by variables such as the density $\rho$ or the conserved rest-mass density $D$. This becomes problematic in regions that would otherwise be vacuum, since these quantities approach zero. To avoid such issues, it is common practice to impose a low-density atmosphere around the star. Although this added material is not physical, it is generally chosen to be sufficiently dilute so that its influence on the system is expected to remain minimal. Despite this assumption, the presence of this atmosphere can alter the evolution of the system and, in some cases, affect the numerical stability of the computation. The flooring scheme adopted by \AK{} artificially injects mass and heat into the fluid and removes momentum from the system when a density threshold is reached. This introduces a small violation of mass conservation in the overall simulation, which we aim to maintain at the level of $10^{-8}$. Such accuracy is generally considered reasonable for most numerical simulations.

A limitation of our simulations arises from the strong surface magnetic fields of the NS, which naturally lead to highly magnetized regions in the low-density atmosphere near the stellar surface, particularly at high resolution. In this regime, the GRMHD equations become increasingly ill-conditioned, which can lead to numerical instability. To maintain stability, we impose a magnetization ceiling such that when a threshold value of $\sigma_{max} = 10^{8}$ is exceeded, the density is increased locally, effectively limiting the magnetization. This procedure also introduces a small additional violation of mass conservation, which remains at the level shown in Fig.~\ref{fig:diag_conv}. At higher resolution the cumulative error becomes non-negligible, as shown for the $\Delta x = 115\, \rm{m}$ case. Here the initial $\nabla \cdot \mathbf{B} / B$ is greater as the sharper field gradients near the stellar surface lead to increased truncation error in the divergence.
A possible approach to mitigate the large magnetization values arising near the stellar surface is to replace the uniform low density medium with a denser atmosphere that decays away from the NS surface. Although this treatment is unphysical, such prescription could be adopted in future numerical simulations to improve numerical stability in strongly magnetized regions.

\bibliographystyle{apsrev4-2}

\end{document}